\documentclass[12pt,dvips]{article}
\usepackage[dvips]{epsfig}
\usepackage{amsmath}

\begin{document}

\title{Free energy of an SU(2) monopole-antimonopole pair.}
\author{Ch.~Hoelbling, C.~Rebbi \\
{\small    Boston University Physics Department}\\
{\small    590 Commonwealth Avenue}\\
{\small    Boston MA 02215, USA}
        \and
V.~A.~Rubakov \\
{\small  Institute for Nuclear Research
  of the Russian Academy of
  Sciences}\\
{\small 60th October Anniversary Prospect 7a}\\
{\small Moscow 117312, Russian Federation}
}

\maketitle

\newcommand{\bp}{{\beta^\prime}}
\newcommand{\Tr}{\mbox{Tr}}

\begin{abstract}
We present a high-statistic numerical study of the free energy of a
monopole-antimonopole pair in pure SU(2) theory. We find that the
monopole-antimonopole interaction potential exhibits a screened
behavior, as one would expect in presence of a monopole condensate.
Screening occurs both in the low-temperature, confining phase of the
theory, and in the high-temperature deconfined phase, with no evidence
of a discontinuity of the screening mass across the transition.  The
mass of the object responsible for the screening at low temperature is
approximately twice the established value for the lightest glueball, 
indicating a prevalent coupling to glueball excitations.  At
high temperature, the screening mass increases.  We contrast the
behavior of the quantum system with that of the corresponding
classical system, where the monopole-antimonopole potential is of the
Coulomb type.
\end{abstract}

\section{Introduction.}

It is well known that some Higgs theories with non-Abelian gauge group
admit stable monopole solutions~\cite{tH74,Po74}.  In certain
cases, most notably in grand unified theories, the residual unbroken
gauge group is non-Abelian.  It is then particularly interesting to
study the interaction between two monopoles, or between monopole and
antimonopole, induced by the quantum fluctuations of the unbroken
gauge group.  Beyond the relevance that these interactions may have
for the original theory, they can help clarify the low energy
properties of the residual theory itself: from the point of view of
the low energy theory, monopoles are point-like external sources of
non-Abelian gauge field, so they act as non-trivial probes of the
strong coupling dynamics.  Indeed, already some time ago
t'Hooft~\cite{tH76} and Mandelstam~\cite{Ma76} proposed that
condensation of magnetic monopoles could be responsible for the
confinement mechanism~\cite{KH92}. 
If the vacuum state of a non-Abelian gauge theory is
characterized by the presence of a monopole condensate, this will
screen the monopole-antimonopole interaction, which should therefore
exhibit a Yukawa-like behavior.  If there is no condensate whatever,
one should expect instead a Coulombic interaction between the
monopoles, as is the case in classical SU(2) theory. Finally, in a
theory characterized by a condensation of electric charges, the
monopole-antimonopole interaction energy should increase linearly with
separation.  While a substantial of work has already been done to
understand the role of monopole condensation for the confinement
mechanism~\cite{KT98,KT99,Co98,Co98a,AFE99,BFGO99,En98}, to the best
of our knowledge a precise determination of the monopole-antimonopole
interaction potential in a quantum non-Abelian theory is still
lacking.  In this paper we plan to fill this void, presenting a
numerical calculation of the monopole-antimonopole potential in the
$SU(2)$ theory.  Our results show that the monopole-antimonopole
interaction potential is screened, buttressing the conjecture of a
monopole condensate.  We also study the monopole-antimonopole system
in the high-temperature, deconfined phase of SU(2).  The interaction
still exhibits screening, which can be ascribed to a magnetic mass.
We will investigate the temperature dependence of this mass.

\section{Monopole-antimonopole configuration and calculation
of the free energy.}
\label{sec:free}

\begin{figure}
\begin{center}
\epsfig{file=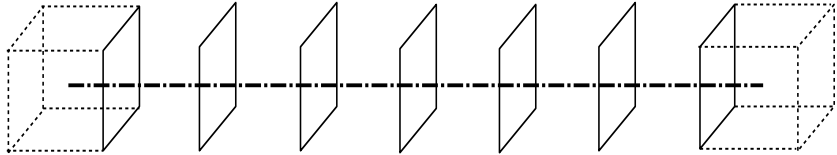,width=11cm,height=2cm}
\end{center}
\caption{\label{chain}Monopole antimonopole pair induced by
twisting the plaquettes transversed by the string.}
\end{figure}

\begin{figure}
\begin{center}
\epsfig{file=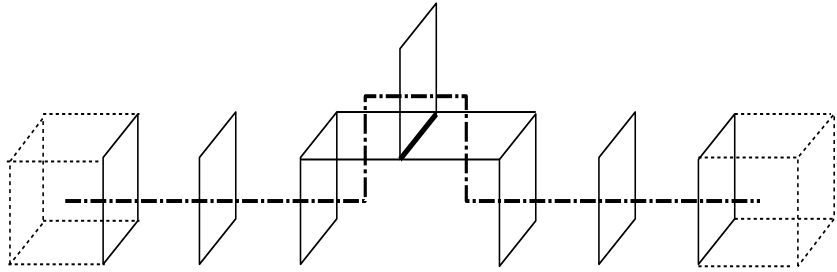,width=11cm,height=3.5cm}
\end{center}
\caption{\label{chain2}Change of the string location brought about by the
redefinition of the gauge variable on the link denoted by a bolder line.}
\end{figure}

The procedure for introducing $SU(N)$ monopole sources on the lattice
was devised by Ukawa, Windey and Guth~\cite{UWG80} and Srednicki and
Susskind~\cite{SS81}, who built on earlier seminal results by
t'Hooft~\cite{tH78}, Mack and Petkova~\cite{MP79,MP80} and
Yaffe~\cite{Ya79}. In this paper we will follow the method of
Ref.~\cite{SS81}.  In three dimensions, an external
monopole-antimonopole pair can be introduced by ``twisting''
the plaquettes transversed by a string joining the monopole 
and the antimonopole (see Fig.~\ref{chain}), i.e.~by changing
the coupling constant of these plaquettes according to
$\beta$ to $z_n \beta$, where $z_n=\exp(2\pi \imath n /N)$ is an
element of the center of the group \footnote{The use of a 
string of plaquettes with modified
couplings to study the disorder was also advocated by Groeneveld, Jurkiewicz 
and Korthals Altes~\cite{GJKA80}.}.  
The location of the string is unphysical.  
It can be changed by redefining link variables on the plaquettes 
transversed by the string by $U \to z_n^{-1} U$, as illustrated in
Fig.~\ref{chain2}.  The fact that $z_n$ is an element of the center
$SU(N)$ guarantees that the above redefinition is a legitimate change
of variables.  On the other hand, the position of the cubes that
terminate the string cannot be changed.  Those cubes contain two
external monopoles of charge $z_n$ and $z_n^{-1}$ respectively (or,
equivalently, a monopole of charge $z_n$ and its antimonopole).  In
the $SU(2)$ theory we consider in this paper, the only non-trivial
element of the center of the group is $z=-1$ and thus monopole and
antimonopole coincide.

In four dimensions, a static monopole-antimonopole pair is induced by
replicating the string on all time slices.  The insertion of a
monopole-antimono-pole pair can be reinterpreted in terms of the
electric flux operator and several investigations have been devoted to
the study of such operator (see for instance
Refs.~\cite{BLS81,DGT82}).  However, as we already stated, accurate
numerical information on the free energy of a monopole-antimonopole
has never been obtained.

The numerical calculation of a free energy by stochastic simulation
can be a quite challenging problem, since the free energy is related
to the partition function and the partition function, which is the
normalizing factor for the simulation, cannot be directly measured.
In our case, we need the ratio of the partition functions of a system
containing the monopole-antimonopole pair to the partition function of
the free system.  In order to calculate this quantity we define a
generalized system where the coupling constant $\beta$ has been
replaced by $\bp$ on all the plaquettes transversed by the string.
For sake of precision, if we place the monopole at the spatial
location of integer value coordinates $x_0+\frac{1}{2},
y_0+\frac{1}{2}, z_0+\frac{1}{2}$ and the antimonopole at
$x_0+\frac{1}{2}, y_0+\frac{1}{2}, z_0+d+\frac{1}{2}$, the plaquettes
with coupling $\bp$ will be all the $x$-$y$ plaquettes with lower
corner in $x_0, y_0, z, t$, where $z_0+1\le z \le z_0+d$ and $0 \le t
\le N_t-1$.  $N_x, N_y, N_z, N_t$ denote the extents of the lattice in
the four dimensions.  The $\frac{1}{2}$ offsets in the coordinates of
the monopole and antimonopole are due to the fact that we consider
them located at the center of two spatial cubes, namely those with
lowest corners in $x_0, y_0, z_0$ and $x_0, y_0, z_0+d$,
respectively. For the monopole-antimonopole configuration we use
periodic boundary conditions.  We will also consider single monopole
configurations.  For these we use periodic boundary conditions only in
time and the two spatial directions orthogonal to the string, while we
choose free boundary conditions in the spatial direction parallel to
the direction of the string.  We let the string run from the mid-point
of the lattice to the free boundary.  This places a single monopole in
the middle of the lattice and emulates as well as possible a
configuration where the antimonopole has been removed to infinity.

Let us denote by $M$ the set of all the plaquettes with modified
coupling constant.  The action of the modified system is then
\begin{equation}
S(\beta,\bp)=\frac{1}{2}\left(\beta\sum_{P\not\in M}\Tr(U_P)+
\bp\sum_{P\in M}\Tr(U_P)\right)
\end{equation}
and the corresponding partition function is
\begin{equation}
\label{modified}
Z(\beta,\bp)=\sum_{\mathcal C} e^{-S(\beta,\bp)}
\end{equation}
where ${\mathcal C}$ denotes the set of all configurations.  For $\bp
= \beta$ our generalized system obviously reduces to a homogeneous
$SU(2)$ lattice gauge system with coupling constant $\beta$, whereas
for $\bp=-\beta$ it becomes an $SU(2)$ model with a static
monopole-antimonopole pair.  In this latter case, as discussed above,
the partition function becomes independent on the actual location of
the string, and depends only on the distance $d$ between monopole and
antimonopole (beyond depending, of course, on $\beta$ and the extent
of the lattice).  The change of free-energy induced by the presence of
the pair is thus
\begin{equation}
\label{freen}
F=-T\log\frac{Z(\beta,-\beta)}{Z(\beta,\beta)}
\end{equation}
where $T=N_t a$ is the temperature of the system ($a$ denotes, as
usual, the lattice spacing).  This is the quantity we want to
calculate.  For $\bp$ not equal to $\beta\ \rm{or}\ -\beta$,
$Z(\beta,\bp)$ does depend on the location of the string.
Nevertheless Eq.~\ref{modified} continues to define the partition
function of a statistical system, which we will use for our
calculation of $F$.  The basic idea is that we will perform Monte
Carlo simulations of the modified system for a set of values of $\bp$
ranging from $\beta$ to $-\beta$ which is sufficiently dense that for
all steps in $\bp$ we can reliably estimate the change induced in
$\log Z$.  Conceptually this corresponds to the observation that for
an infinitesimal change in $\bp$ the change in free energy will be
\begin{equation}
\frac{\partial F_\bp}{\partial\bp}=
\frac{1}{2}\langle\sum_{P\in M}\Tr(U_P) \rangle_\bp
\end{equation}
and thus the free energy of the monopole pair can be computed as
\begin{equation}
\label{integral}
F=\int_{-\beta}^\beta d\bp\,\frac{\partial F_\bp}{\partial\bp}=
\frac{1}{2}\int_{-\beta}^\beta d\bp\,\langle\sum_{P\in M}\Tr(U_P) \rangle_\bp
\end{equation}
where the integrand in the r.h.s.~is an observable, namely the energy
\begin{equation}
\label{energy}
E=\frac{1}{2}\langle\sum_{P\in M}\Tr(U_P) \rangle_\bp
\end{equation}
of the strip of plaquettes transversed by the string.  However,
implementing Eq.~\ref{integral} would be inefficient, due to
the large number of subdivisions that the numerical integration would
require for accuracy.  Rather, we calculate $F$ by following the
Ferrenberg-Swendsen multi-histogram method~\cite{FS}.

Following Ref.~\cite{FS}, we choose a set of $N+1$ values $\{\bp_i\}$
ranging from $\bp_0=\beta$ to $\bp_N=-\beta$. In the actual
calculation, we chose them equally spaced, although this is not
necessary.  The number $N$ is determined by a criterion that will be
explained below.  For each value $\bp_i$ we perform a Monte Carlo
simulation of the corresponding system and record the values of the
energy $E$ of the plaquettes transversed by the string (see
Eq.~\ref{energy}) in a histogram $h_i(E)$.  More exactly, in our
calculation rather than simply accumulating the entries in the
histograms, we distributed the measured energy values over the four
neighboring vertices according to the weights of a cubic interpolation
formula.  This substantially increases the accuracy when the values in
the histograms are subsequently used to approximate integrals over the
density of states: $\int \rho(E) f(E) dE \approx Z \sum_E (h(E)/n)
f(E)$, $n$ being the total number of entries in the histogram.
Indeed, by using this procedure, we were able to reduce the number of
bins to $200$ without noticeable discretization effects.

 From each separate histogram one can obtain an independent estimate
of the density of states
\begin{equation}
\label{rhoi}
\frac{\rho_i(E)}{Z_i}=\frac{h_i(E)}{n_i}e^{\bp_i E}
\end{equation}
where $Z_i$ is the partition function for the specific value $\bp_i$
and $n_i$ is the total number of histogram entries.  Of course, the
normalizing factors $Z_i$ are still not known.  Indeed, the entire
goal of the computation is to calculate the relative magnitude of the
partition functions $Z_i$.  Starting from Eqs.~\ref{rhoi} one can
however obtain the partition functions $Z_i$ up to a common constant
of proportionality by a self-consistent procedure.  We start from the
crude approximation that all $Z_i$ are equal.  Since we are only
interested in ratios of partition functions, we can set this common
value to 1.  We combine then the estimates of the density of states
given by Eqs.~\ref{rhoi} into a first approximation
\begin{equation}
\rho(E)=\frac{1}{N}\sum_i\rho_i(E)
\end{equation}
 From this value of the density of states we can now obtain a better
approximation of the partition functions
\begin{equation}
\label{partf}
Z_i=\sum_E \rho(E) e^{-\bp_i E}
\end{equation}
Equations~(\ref{rhoi}-\ref{partf}) can now be iterated to get $Z_i$ up
to a multiplicative factor. If the iterations converge, the final
results will provide a self-consistent set of values for the partition
functions $Z_i$, up to a common constant.  In particular, we will be
able to obtain $\frac{Z(\beta,-\beta)}{Z(\beta,\beta)}=
\frac{Z_N}{Z_0}$ and the free energy of the monopole-antimonopole pair.

\begin{figure}
\begin{center}
\epsfig{file=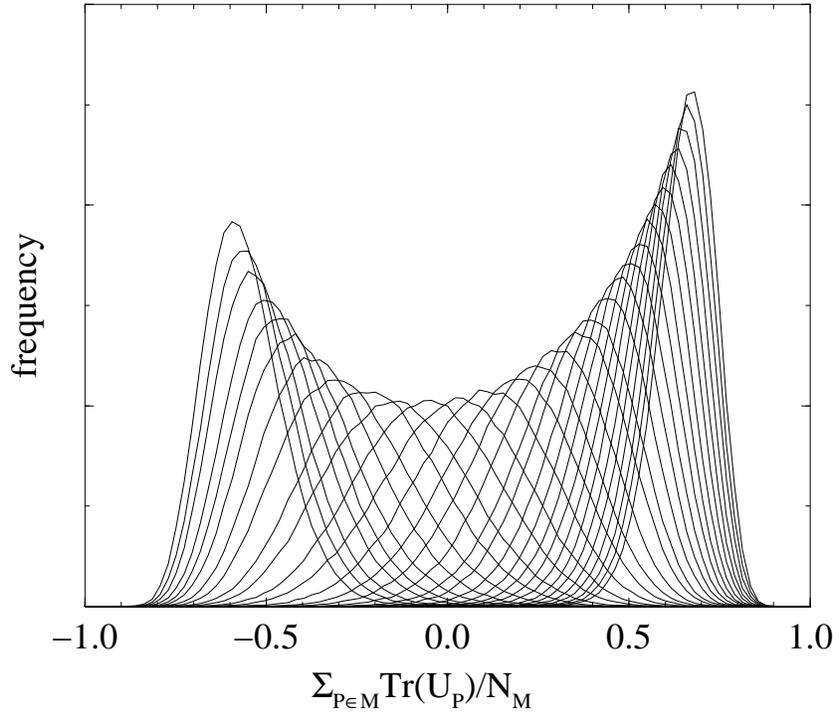,width=11cm}
\end{center}
\caption{\label{multihisto}The overlap of histograms at $\beta=2.6$ and
different $\beta^\prime$. The monopole separation is $2a$ and $N_t=6$.}
\end{figure}

A necessary condition for the above procedure to work well is that the
histograms corresponding to adjacent $\bp$ have a sufficient overlap.
In fig.~\ref{multihisto} we plot the histograms obtained in a typical
run and one can see that they exhibit indeed a substantial overlap.
In our calculation we found that the self-consistent procedure outlined
above converged rapidly for all the values of lattice size, coupling
constant and monopole-antimonopole separation we considered.  We checked
that the results for the free energy we obtained with the multihistogram 
method are consistent with the values one can obtain from the numerical 
integration of $\partial F_{\beta'}/\partial \beta'$ 
(cfr.~Eq.~\ref{integral}), within the error of the latter procedure.

\section{Computational details and numerical results.}

We simulated a pure $SU(2)$ system with a combined overrelaxation and
multihit-Metropolis algorithm for minimal autocorrelation time.  We
studied systems with coupling constants varying between $\beta=2.476$
and $\beta=2.82$, spatial sizes ranging from $16^2 \times 32$ to $32^2
\times 64$ and Euclidean temporal extents ranging from $16$ down to
$2$. The corresponding temperatures span the deconfinement phase
transition, which occurs at $N_t\approx 10$ for $\beta=2.6$ and
between $N_t=7$ and $N_t=8$ for $\beta=2.476$ and
$\beta=2.5$~\cite{FHK92}.  Full details of lattice sizes and couplings
used in our simulations are given in table \ref{runs}.  For each
lattice size, value of $\beta$ and monopole separation, we performed a
sequence of simulations, starting with $\bp=\beta$ and decreasing
$\bp$ in steps (cfr.~Sect.~\ref{sec:free}) to its final value
$\bp=-\beta$.  Precisely, we performed $5000$ thermalization steps at
$\bp=\beta$ followed by measurements separated by $50$ updates. We
decreased $\bp$, performed another $500$ thermalization followed by
the same number of measurements, again separated by $50$ updates, and
so on, until completion of the measurements with $\bp=-\beta$.  For
each measurement, as a variance reduction technique, we performed
$384$ upgrades of the links in the plaquettes belonging to the flux
tube $M$, while keeping all other link variables fixed. In this way,
we obtained $384$ histogram entries per configuration.

\begin{table}[h!]
\begin{center}
\begin{tabular}{r|l|l}
$\beta$ & $N_x\times N_y\times N_z$ & $N_t$ \\
\hline
2.82 & $32\times 32\times 64$ & 4 \\
2.6 & $20\times 20\times 40$ & 2,4,6,16 \\
2.5 & $16\times 16\times 32$ & 4,12 \\
2.476 & $16\times 16\times 32$ & 4,12
\end{tabular}
\end{center}
\caption{\label{runs}Lattice sizes and couplings at which the simulation was
performed.}
\end{table}

The number of measurements in each individual simulation, as well as
the number of steps in $\bp$, depended on monopole separation and
lattice size and are given in tables \ref{dets1} and \ref{dets2}.  
In order to estimate the error we proceeded as follows. For all
data we performed a standard jackknife evaluation of the error based
on 10 subsamples.  The data are however highly correlated and this
leads to an underestimate of the error.  An error analysis based
on the full correlation matrix would have been computationally
too costly.  Instead, we performed seven totally independent calculations
of the free energy for a few data points and calculated the error from
the variance of the results.  This came out approximately four times
larger than the corresponding estimate of the error from the jackknife
method.  Thus we multiplied all of the jackknife errors by a factor
of four.  While this universal rescaling can only produce an approximation 
to the true errors, because the correlation in the data will generally 
vary from data point to data point, we feel that it gives the most 
realistic estimate of the actual errors that can be obtained without 
embarking in an error analysis of prohibitive cost.  Our code was 
written in Fortran 90 and was run on the SGI-Origin 2000 at the 
Boston University Center for Computational Science. It performs 
at $\approx 140$ MFlops on a single 190MHz R10000 CPU and scales 
well up to 64 CPU's. The total CPU-time needed for the simulations 
was $\approx 3\times 10^4$ CPU hours.

\bigskip
\begin{table}[h!]
\begin{center}
\begin{tabular}{r|l|l||l|l||l|l||l|l}
$d$ & \# $\bp$ & \# m & \#$\bp$ & \# m & \#$\bp$ & \# m & \#$\bp$ & \# m  \\
\hline
1 & 11 & 200 & 11 & 600 & 11 & 600 & 11 & 800 \\
2 & 21 & 200 & 21 & 600 & 21 & 600 & 21 & 800 \\
3 & 31 & 200 & 31 & 600 & 31 & 600 & 31 & 800 \\
4 & 41 & 200 & 41 & 600 & 41 & 600 & 41 & 800 \\
5 & 51 & 200 & 51 & 600 &    &     & 51 & 600 \\
6 & 51 & 200 & 61 & 600 & 61 & 600 & 61 & 800 \\
$\infty$ & 101 & 200 & 101 & 600 & & & &
\end{tabular}
\end{center}
\caption{\label{dets1}Computational details of the simulations at $\beta=2.6$,
and $N_t=16,\; 6,\; 4,\; 2$ (from left to right). $d$ is the monopole
separation in lattice units, \# $\bp$ is the number of $\bp$ steps
and \# m is the number of independent configurations per $\bp$,
which measurements were taken over.}
\end{table}

\begin{table}[h!]
\begin{center}
\begin{tabular}{r|l|l||l|l||l|l||l|l||l|l}
$d$ & \# $\bp$ & \# m & \#$\bp$ & \# m & \#$\bp$ & \# m & \#$\bp$ & \# m &
\#$\bp$ & \# m  \\
\hline
1 & 11 & 100 & 11 & 600 & 11 & 600 & 11 & 200 & 11 & 800 \\
2 & 21 & 100 & 21 & 600 & 21 & 600 & 21 & 200 & 21 & 800 \\
3 & 31 & 100 & 31 & 600 & 31 & 600 & 31 & 200 & 31 & 800 \\
4 & 41 & 100 & 41 & 600 & 41 & 600 & 41 & 200 & 41 & 800 \\
6 & 61 & 100 & 61 & 600 & 61 & 600 & 61 & 200 & 61 & 800 \\
\end{tabular}
\end{center}
\caption{\label{dets2}Computational details of the simulations
at $\beta=2.82,\; N_t=4$; $\beta=2.5,\; N_t=12$; $\beta=2.5,\; N_t=4$;
$\beta=2.476,\; N_t=12$ and $\beta=2.476,\; N_t=4$ (from left to right).}
\end{table}

\vfil\eject
We measured the free energy of the monopole-antimonopole pair
for several values of lattice size, coupling constant $\beta$
and monopole-antimonopole separation $d$. Tables \ref{result26} and
\ref{result25} list all our results.

\begin{table}
\begin{center}
\begin{tabular}{r|l|l|l|l}
$d$ & $F[N_t=16]$ & $F[N_t=6]$ & $F[N_t=4]$ & $F[N_t=2]$ \\
\hline
1 & 1.3464(50)& 1.3472(38)& 1.3193(92)& 1.0236(114) \\
2 & 1.6961(28)& 1.6741(48)& 1.5811(51)& 1.1492(98) \\
3 & 1.7708(18)& 1.7357(20)& 1.6056(35)& 1.1586(69) \\
4 & 1.7909(18)& 1.7334(25)& 1.6253(34)& 1.1531(62) \\
5 & 1.7927(10)& 1.7350(19)& & \\
6 & 1.7950(10)& 1.7375(33)& 1.6411(26)& 1.1420(44) \\
$\infty$ & 1.7927(22)& & &
\end{tabular}
\end{center}
\caption{\label{result26}Free energy in units of $a^{-1}$
of the monopole-antimonopole pair at various separations $d$ for the
confined ($N_t=16$) and deconfined ($N_t=6,4,2$) phases.  Results are
for spatial lattice size $N_x\times N_y\times N_z=20\times 20\times
40$ at $\beta=2.6$. The entry with separation $\infty$ is twice the
energy of a single monopole as measured on a $N_x\times N_y\times
N_z=20\times 20\times 20$ lattice and free boundary conditions in the
$z$ direction.}
\end{table}

\begin{table}
\begin{center}
\begin{tabular}{r|l||l|l||l|l}
$d$ & $F[N_t=4]$ & $F[N_t=12]$ & $F[N_t=4]$ & $F[N_t=12]$ & $F[N_t=4]$ \\
\hline
1 & 1.648(40) & 1.1451(58) & 1.1332(92) & 1.0988(35) & 1.0927(42)\\
2 & 2.037(28) & 1.4020(33) & 1.3549(41) & 1.3241(24) & 1.2885(53)\\
3 & 2.129(20) & 1.4437(28) & 1.3741(36) & 1.3574(16) & 1.3133(40)\\
4 & 2.222(16) & 1.4601(24) & 1.3806(37) & 1.3736(14) & 1.3188(26)\\
6 & 2.191(17) & 1.4726(23) & 1.3779(26) & 1.3683(26) & 1.3169(18)
\end{tabular}
\end{center}
\caption{\label{result25}Free energy in units of $a^{-1}$
of the monopole-antimonopole pair at various separations $d$ at high
temperature ($N_t=4$) on a $N_x\times N_y\times N_z=32\times 32\times
64$ lattice at $\beta=2.82$ (left), and for the confined ($N_t=12$)
and deconfined ($N_t=4$) phases on a $N_x\times N_y\times N_z=16\times
16\times 32$ lattice at $\beta=2.5$ (middle) and $\beta=2.476$
(right).}
\end{table}

\vfil\eject

\begin{figure}[h!]
\begin{center}
\epsfig{file=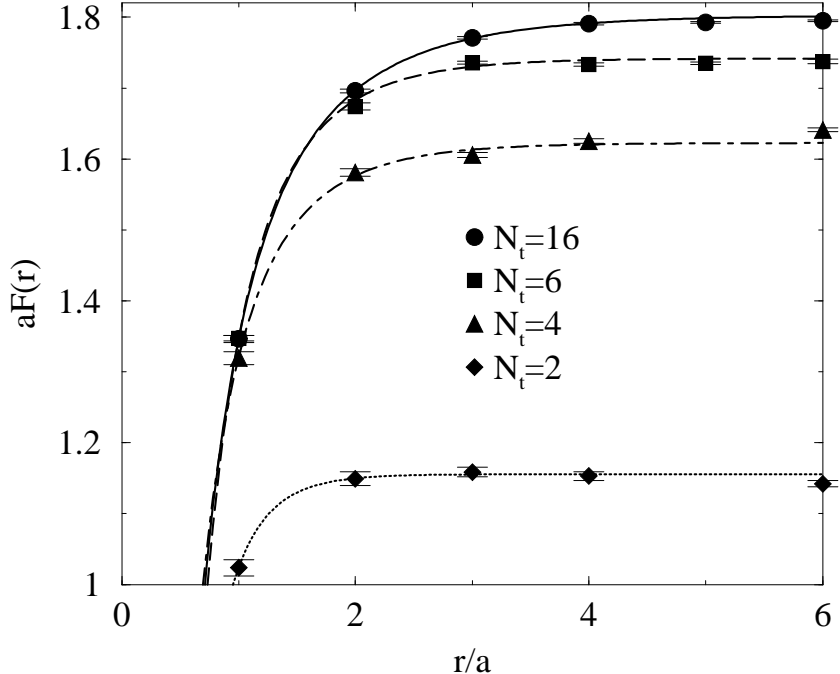,width=11cm}
\end{center}
\caption{\label{fmass} Free energy of the monopole-antimonopole pair
versus separation at $\beta=2.6$ for different values of $N_t$. The lines
represent fits of the first 4 points to a Yukawa potential. The
results of the fit for the screening mass $m$ are listed in
table~\ref{tmass}.}
\end{figure}

Typical results are illustrated in Figure~\ref{fmass}, where we plot
the data for the monopole-antimonopole free energy obtained at
$\beta=2.6$.  The flattening of the potential at large separation
gives evidence for the screening of the interaction.  The lines in the
figure reproduce fits to a Yukawa potential
\begin{equation}
F(r)=F_0-c\frac{e^{-mr}}{r}
\label{yukawa}
\end{equation}
The fits give clear indications for a non-vanishing screening mass
(see the tables for details) and rule out a Coulombic behavior
of the potential.

We fit all of our data to a Yukawa potential, as in Eq.~\ref{yukawa}.  We
used the points at separation $d=1$ to $4$ for the fits.  In the confined
phase, it is also possible to perform meaningful fits through the points at
separation $2$ to $6$, leaving out the point at $d=1$, where one expects the
value of the potential to be most affected by lattice distortions.  For
higher temperatures the rapid flattening of the potential makes the fits more
sensitive to the removal of the first point.  An alternative procedure
consists in fitting the data to a lattice Yukawa potential\footnote{We are
  grateful to M.~Chernodub and M.~Polikarpov for bringing this point to our
  attention.}, as suggested in this context in Ref.~\cite{ch00}. The results
of the fits for the screening masses are reproduced in Table~\ref{tmass}.
Unprimed (primed) quantities refer to the values obtained from fits to a
continuum (lattice) potential of the data with $d$ ranging from $1$ to $4$.
For the conversion in units of the string tension $\sqrt{\sigma}$ we used
$a\sqrt{\sigma}=0.1989,0.1834, 0.1326,0.0663$ for $\beta=2.476,2.5,2.6,2.82$,
respectively.  We took the values for $\beta=2.5,2.6$ from Ref.~\cite{Te98}
and calculated the other values from the known scaling behavior of the
theory.  The fact that all fits produce a non-vanishing value for the
screening mass is a clear indication that the data are not consistent with a
Coulombic monopole-antimonopole interaction.  To reinforce this point, we
attempted a direct Coulombic fit to the data for $\beta=2.6$, $N_t=16,\; 6$
and found the fit quality to be $Q\approx 10^{-62}$ for $N_t=16$ and $Q\approx
10^{-68}$ for $N_t=6$, definitely ruling out a Coulombic behavior of the
potential in both phases.

\begin{table}
\begin{center}
\begin{tabular}{r|l|l|l|l|l|l|l}
$\beta$ & $T/\sqrt{\sigma}$ & $m/\sqrt{\sigma}$ & $ma$ & Q & 
$m'/\sqrt{\sigma}$ & $m'a$ & Q' \\
\hline
2.476 & 0.419 & 4.62(32) & 0.918(64) & $0.003$ & 5.87(45) & 
1.168(89) & $0.0001$\\
2.5 & 0.454 & 5.04(43) & 0.924(78) & 0.16 & 6.43(60) & 1.180(110) & 0.06 \\
2.6 &  0.472 & 5.79(31) & 0.768(41) & 0.54 & 7.21(40) & 0.956(53) & 0.67
\end{tabular}

\vskip 4mm

\begin{tabular}{r|l|l|l|l|l|l|l}
$\beta$ & $T/\sqrt{\sigma}$ & $m/\sqrt{\sigma}$ & $ma$ & Q & 
$m'/\sqrt{\sigma}$ & $m'a$ & Q' \\
\hline
2.476 & 1.257 & 6.4(1.0) & 1.28(20) & 0.85 & 8.6(1.7) & 1.71(34) & 0.72 \\
2.5 & 1.363 & 8.65(1.29) & 1.59(24) & 0.47 & 12.49(2.34) & 2.29(43) & 0.40 \\
2.6 & 1.257 & 9.16(69) & 1.22(9) & $0.0001$ & 11.86(1.11) 
& 1.57(15) & $0.0004$
\end{tabular}

\vskip 4mm

\begin{tabular}{r|l|l|l|l|l|l|l|l}
$\beta$ & $N_t$ & $T/\sqrt{\sigma}$ & $m/\sqrt{\sigma}$ & $ma$ & $Q$ & 
$m'/\sqrt{\sigma}$ & $m'a$ & Q' \\
\hline
2.6 & 4 & 1.885 & 8.9(1.2) & 1.18(16) & $0.006$ & 12.0(2.0) 
& 1.59(26) & $0.003$ \\
2.6 & 2 & 3.771 & 18.5(8) & 2.45(11) & 0.53 & 31.3(17.3) & 4.15(2.30) & 0.53 \\
2.82 & 4 & 3.771 & 7.9(2.6) & 0.52(17) & 0.08 & 8.4(3.0) & 0.55(20) & 0.12
\end{tabular}
\end{center}
\caption{\label{tmass}Screening masses from a Yukawa fit of the free energy
in units of the zero-temperature string tension $\sqrt{\sigma}$ and in
lattice units. For comparison, the temperature of the system in units of
the zero-temperature string tension is also given. The first table
is for confined systems at at comparable temperatures.  The second table
is for deconfined systems, again at comparable temperatures.  The third
table gives the screening masses for various high temperature simulations.
$m$ and $m'$ refer to the fits done with continuum and lattice potential, 
respectively. $Q$ and $Q'$ are the corresponding qualities of the fit.}
\end{table}

In table \ref{tmass2} we compare our values for the screening mass at
$\beta=2.5$ and $\beta=2.6$ with with the values for the lightest
glueball mass in the SU(2) theory~\cite{Te98,MT88}.  We reproduce the
results from fits with the continuum Yukawa potential to the data
points at separation $1$ through $6$ ($m_1$), from fits, always with
the continuum potential, where we disregarded the points at separation
$1$ which have the largest discretization error ($m_2$), and from
fits to the points at separation $1$ through $6$ with the lattice
Yukawa potential ($m'$).  The values $m_2$ and $m'$ are consistent 
and are approximately twice as large as the mass of the lightest
glueball.  This indicates that the predominant coupling of the
monopoles is to glueball excitations.  Our results do not rule
out that the lightest glueball may dominate screening at long
distances, but this not visible within the range of lattice
separations ($a - 5a$) for which we can obtain sufficiently
accurate results.

\begin{table}[h!]
\begin{center}
\begin{tabular}{r|l|l|l|l|l|l|l}
$\beta$ & $m_g a$& $m_1 a$ & $m_2 a$ & $m'a$ & $Q(m_1)$ & $Q(m_2)$ & $Q(m')$ \\
\hline
2.6 & 0.51(3) & 0.849(26) & 1.019(14) & 0.998(36) & 0.08 & 0.52 & 0.59
\end{tabular}
\end{center}
\caption{\label{tmass2}Comparison of screening masses and the lowest
glueball mass $m_g a$ (taken from \cite{Te98,MT88}). The masses $m_1$ 
and $m_2$ are obtained from fits with the continuum Yukawa potential to 
the data points at separation $1$ through $6$ and $2$ through $6$,
respectively. The masses $m'$ are from fits with the lattice Yukawa
potential to the data points at separation $1$ through $6$.}
\end{table}

\begin{figure}[th!]
\begin{center}
\epsfig{file=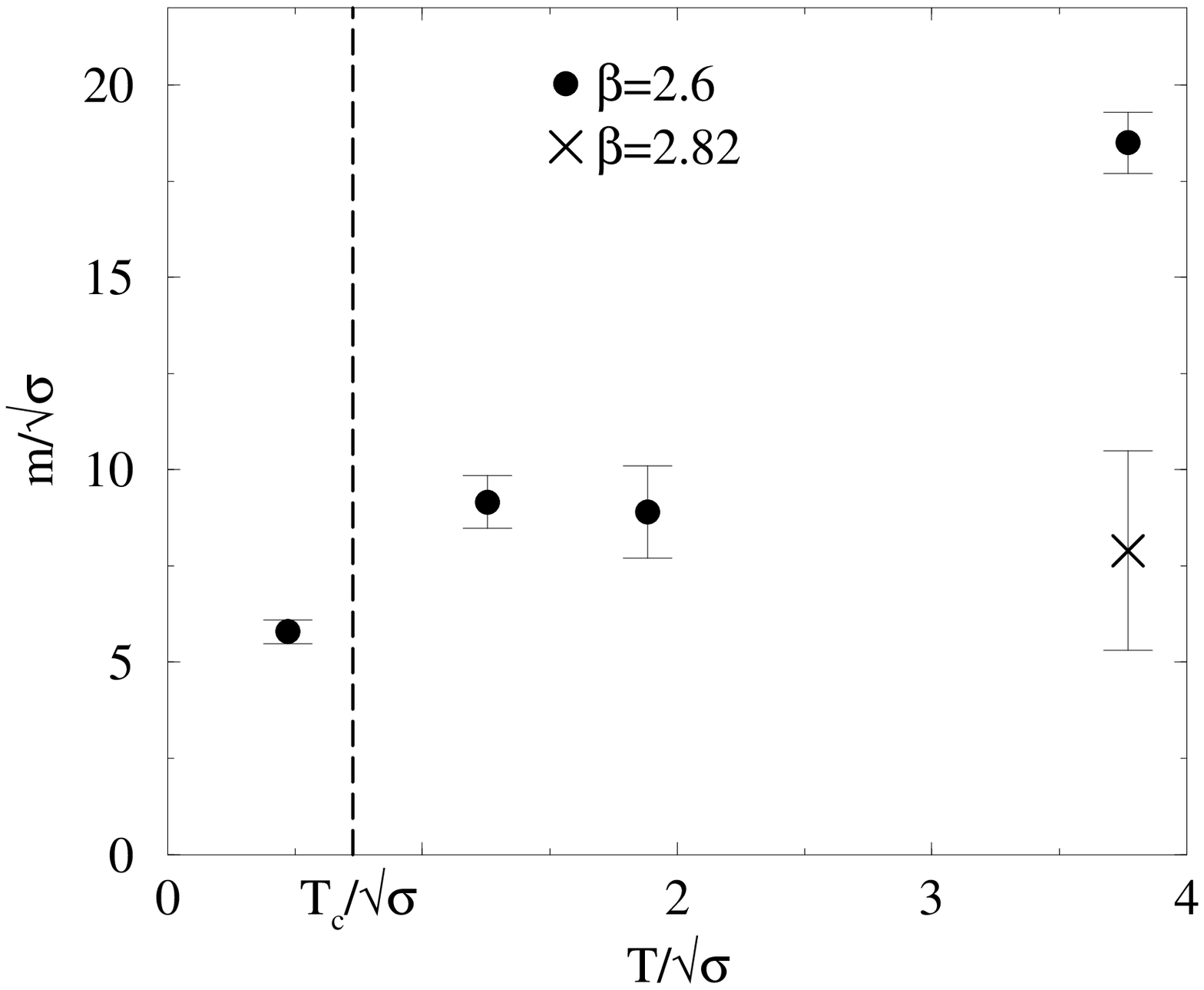,width=11cm}
\end{center}
\caption{\label{fhit}Screening mass vs.~temperature at $\beta=2.6$.
Because of the large systematic errors introduced by the small temporal
lattice extent $N_t=2$ for the last data point, we also included results
from simulation at the same physical volume and half the lattice spacing
($\beta=2.82$). The dashed line indicates the critical temperature
(taken from \cite{FHK92}).}
\end{figure}

\begin{figure}[h!]
\begin{center}
\epsfig{file=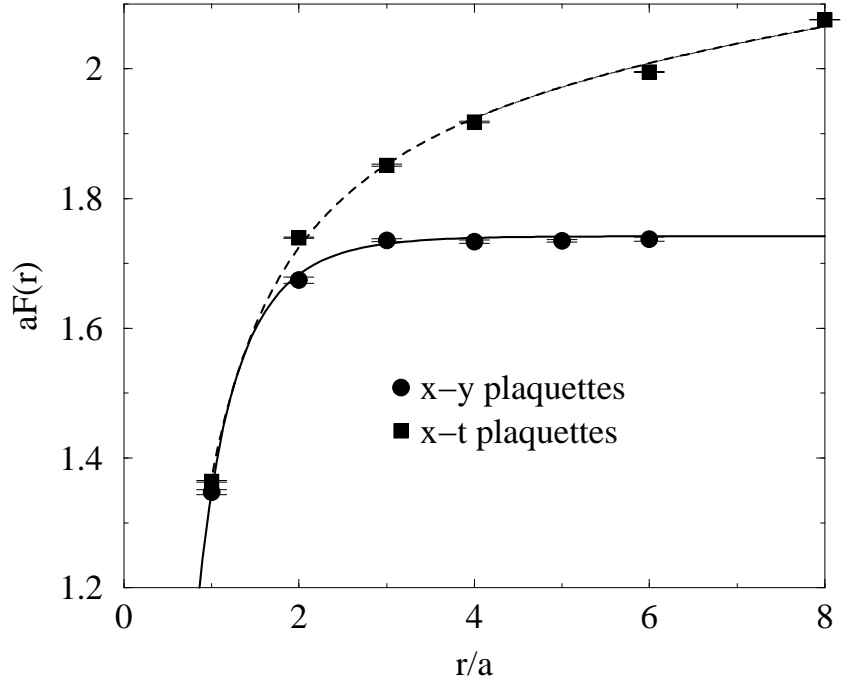,width=11cm}
\end{center}
\caption{\label{f26}Comparison of the ``free energy'' for propagation
in the space direction vs.~propagation in the time direction of the
monopole-antimonopole pair, for a system in the deconfined phase.}
\end{figure}

In figure \ref{fhit} we plot our results for the screening mass $m$
vs.~temperature.  At high temperature $m$ should be identified with
the magnetic screening mass.  Our results are consistent with data
obtained by Stack~\cite{St96} by another method.  They appear
to be somewhat larger than the values for $m$ obtained, by a yet
different technique, by Heller, Karsch and Rank~\cite{HKR98}.  The
authors of Ref.~\cite{HKR98} quote results, however, for systems with
larger $N_t$ and higher $\beta$ than we generally used in our
investigation.  The closest comparison can be made between our
result for $\beta=2.82$, namely $m/T=2.09(0.69)$, and the results
in Ref.~\cite{HKR98}: $m/T=2.01(0.29), 1.24(0.04)$ for $\beta=2.74,
2.88$ respectively.  These latter sets of results are reasonably
consistent. It is interesting that, within the accuracy of our
data, there is no indication for a discontinuous behavior of $m$ at
the deconfinement transition. The apparently continuous behavior of
$m$ should not come, though, as a surprise.  We should remember that
the quantity we study is based on insertion in the SU(2) theory of an
operator (the sheet of plaquettes with modified coupling joining the
world line of the monopoles) which is dual to the $x-y$ Wilson
loops~\cite{tH76}. Space-space Wilson loops exhibit an area law
behavior both below and above the deconfinement transition and,
correspondingly, one would expect that the free energy of
monopole-antimonopole pair, whose propagation spans a dual space-time
surface, should exhibit screened behavior on both sides of the phase
transition.  The discontinuity at the phase transition occurs in the
behavior of space-time Wilson loops or in the correlation of timelike
Polyakov loops.  Accordingly, we would expect a discontinuity in the
partition function of the system with the monopole-antimonopole pair
propagating in the space direction. In order to test this idea, we
also measured the partition function of a system where we changed the
sign of the $t-y$ plaquettes crossing a string joining monopole and
antimonopole separated by $d$ lattice sites in the $z$ direction and
propagating in the $x$ direction.  We performed the calculation at
$\beta=2.6$ with a lattice of size $N_x=N_y=20, N_z=40, N_t=6$.  While
the physical meaning of the ``free energy'' $F=-(1/N_x)
\log[Z(\beta,-\beta)/Z(\beta,\beta)]$ become less obvious (it would be
the free energy for a low-temperature system confined in a periodic
box of width $N_x$), our results, listed in table~\ref{f26t} and
illustrated in Fig.~\ref{f26}, show that above the phase transition
this quantity does exhibit a confined behavior.  The dashed line in 
the figure reproduces a fit of a Coulomb plus linear form
$a+b/x+c x$ with parameters $a=2.107(2), b=0.658(3), c=0.0167(3)$.
It is interesting to observe that from the fit we get 
$\sqrt{c}=0.1291(11)$, while the string tension on a $20^4$ lattice
at the same value of $\beta$ is $a\sqrt{\sigma}=0.1326(30)$.

\begin{table}
\begin{center}
\begin{tabular}{r|l}
$d$ & $F$  \\
\hline
1 & 1.3640(13) \\ 
2 & 1.7396(10) \\
3 & 1.8512(15) \\
4 & 1.9180(13) \\
6 & 1.9948(6)  \\
8 & 2.0756(6)
\end{tabular}
\end{center}
\caption{\label{f26t} Value of the spatial 't Hooft loop in units of
$a^{-1}$ at various separations $d$ for a deconfined system ($N_t=6$)
with spatial lattice size $N_x\times N_y\times N_z=20\times 20\times
40$ at $\beta=2.6$.}
\end{table}

It is interesting to compare our results to the solutions of the
classical theory. Since to the best knowledge of the authors there is no
analytical solution to the classical two monopole problem, we have
investigated its properties numerically. To do this, we found the minimal
energy solutions on a lattice and checked their behavior.  We put the
classical system on a 3-dimensional lattice with free boundary
conditions.  (We used free boundary condition because the calculation
itself shows that the potential has a long range behavior and with
free boundary conditions we can reduce finite size effects.)  We
started both from a random non-Abelian configuration and and from a
random Abelian configuration and performed iterative local
minimization to relax the system to its lowest energy state.  There
were no surprises as we found that for both initial conditions the
system relaxed to a minimal energy state of the same energy and that
the non-Abelian solution, after going to a maximally Abelian gauge,
turned out to be entirely Abelian in nature. Also, the potential of
interaction was well fit by the Coulomb form $V=1/(4r)$. (One expects
a coefficient 1/4 in the Coulomb potential because the total magnetic flux
from the monopole is $\Phi=\pi$. This value has been numerically confirmed
in our calculation.)

\begin{figure}[h!]
\begin{center}
\epsfig{file=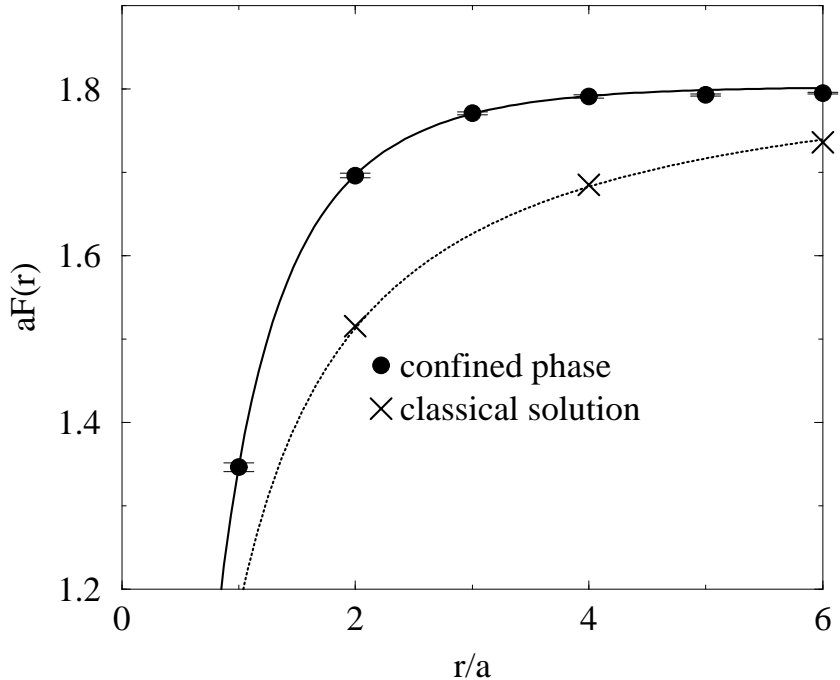,width=11cm}
\end{center}
\caption{\label{compare}Comparison of monopole-antimonopole potentials
in the confined phase of the quantum system and in the classical system.
Data for the quantum system are at $\beta=2.6$ and $N_t=16$.}
\end{figure}

\begin{figure}[h!]
\begin{center}
\epsfig{file=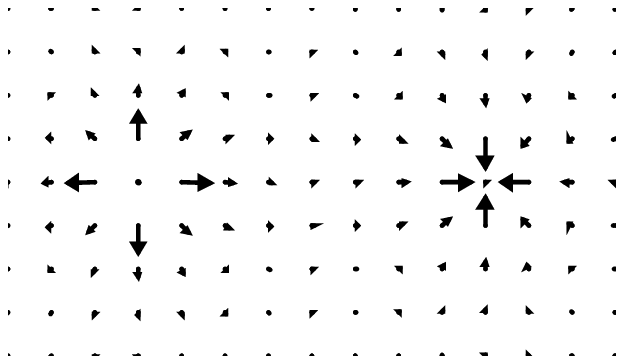,width=7cm}

\epsfig{file=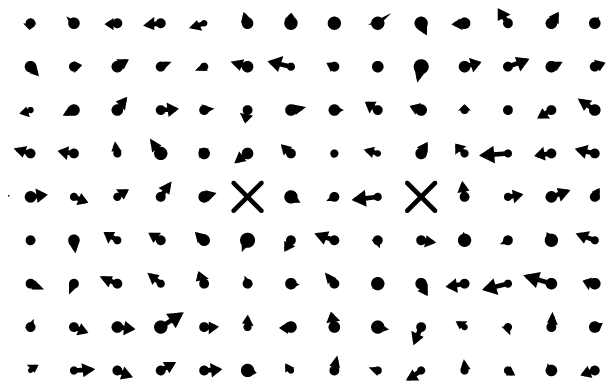,width=7cm}

\epsfig{file=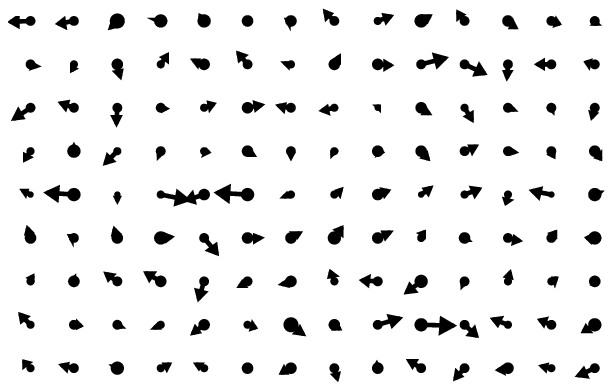,width=7cm}
\end{center}
\caption{\label{pic}Comparison of the maximally Abelian projected
configuration for the classical minimal energy solution (first picture) and
an average over the time slices of a typical quantum configuration. The
second picture is a detail of a region around the monopoles, the third one
is from a region without monopoles.}
\end{figure}

\vfil\eject
In Figure~\ref{compare} we compare the monopole-antimonopole
potentials in the confined phase of the quantum system and in the
classical system.  The comparison shows a clear difference between the
classical and quantum case and reinforces the conclusion that quantum
fluctuations introduce a screening of magnetic monopoles.  We further
illustrate this point by displaying in Fig.~\ref{pic} snapshots of a
typical quantum configuration (from a simulation at $\beta=2.476$ and
$N_t=12$) and of the classical solution.  In the pictures, vectors
show the magnetic field in a maximal Abelian projection and the size
of the dots measure the non-Abelian character of the configuration at
that point (it is proportional to the square of the components
orthogonal to the Abelian projection).  In the classical case the
location of the monopole-antimonopole pair is evident.  In the quantum
case it is marked by the crosses in the middle of the second picture.
Had we not marked the location of monopole and antimonopole in the
quantum case, the reader would be hard pressed in finding where they
are.  The marked difference between the classical and quantum
configurations gives a vivid illustration of how the quantum
fluctuations of the gluon field provide a mechanism for the screening
of external monopole sources, which is most likely also responsible
for confinement in the low temperature phase and for the emergence of
the magnetic mass in the high temperature phase.

\section{Conclusions.}

We have measured the free energy of a monopole-antimonopole pair in
pure SU(2) gauge theory at finite temperature. We find that the
interaction is screened in both the confined and deconfined phase.
The mass of the object responsible for the screening at low
temperature is approximately twice the established value for the
lightest glueball, indicating a prevalent coupling to glueball
excitations.  There is no noticeable discontinuity in the screening
mass at the deconfinement transition, but in the deconfined phase we
clearly see an increase of the screening mass with temperature. Our
results support the hypothesis of existence of monopole condensate in
the vacuum of the SU(2) theory and provide evidence that some glueball 
excitation could serve as a ``dual photon'' in the dual superconductor
hypothesis of quark confinement.  Finally, we would like to observe
that the method we have developed for the calculation of the
monopole-antimonopole free energy is applicable to other models,
beyond the SU(2) theory considered in this paper.  While moderately
demanding in computer resources, it appears capable of producing
accurate numerical results for the monopole-antimonopole potential of
interaction.  Thus it could be used to shed light on the dynamics of
other interesting systems that are expected to exhibit the formation
of electric or magnetic condensates in their vacuum states.

{\bf Acknowledgments.}
We gratefully acknowledge conversations and exchanges of correspondence 
with Maxim Chernodub, Urs Heller, Chris Korthals Altes, Christian Lang, 
Peter Petreczky, Misha Polikarpov, John Stack, Matthew Strassler and 
Terry Tomboulis.  We also grateful to Philippe de Forcrand for bringing 
to our attention a discrepancy between his own results for the 
monopole-antimonopole free energy and the values we presented
in an earlier version of this paper, which helped us correct
a programming error in the implementation of the multihistogram
method, and for correspondence regarding the free energy of
the space-like 't Hooft loop.  This research was supported in 
part under DOE grant DE-FG02-91ER40676 and by the U.S.~Civilian 
Research and Development Foundation for Independent States of 
FSU (CRDF) award RP1-187.

\end{document}